\documentclass[a4paper, usenatbib]{mnras}
\usepackage{graphicx}
\usepackage{epsfig}
\usepackage{subfig}
\usepackage{multirow}
\usepackage{mathrsfs,amsmath} 
\usepackage[utf8]{inputenc}

\def \HI {H~{\sc i}\ }

\title[\HI column density statistics from  absorption]{\HI column  density statistics of the cold neutral medium from absorption studies}

\author[P. K.  Vishwakarma and P. Dutta]{Pavan Kumar Vishwakarma, $^{1}$\thanks{E-mail:pavankv.rs.phy15@itbhu.ac.in} Prasun Dutta,$^{1}$\thanks{E-mail:pdutta.phy@itbhu.ac.in}\\
  $^{1}$ Indian Institute of Technology (BHU), Varanasi, Varanasi 221005, India}
\date{Accepted XXX. Received YYY; in original form ZZZ}
\pubyear{2019}

\begin{document}
\label{firstpage}
\pagerange{\pageref{firstpage}--\pageref{lastpage}}
\maketitle

\begin{abstract}
  Physical properties of the tiny scale structures in the cold  neutral  medium (CNM) of galaxies is a long-standing  puzzle. Only a few lines of sights in our Galaxy have been studies with mixed results on the scale-invariant properties of such structures. Moreover, since these studies measure the variation of neutral hydrogen optical depth, they do not directly constrain the  density structures. In this letter, we investigate the possibility of  measuring the properties  of density and spin temperature structures of  the \HI from  absorption studies of \HI. Our calculations show that irrespective of  the thermal properties of  the  clouds,  the scale structure of the \HI column density can be estimated, whereas, \HI absorption studies alone cannot  shed much light on either the   amplitude of the  density fluctuations and their temperature  structures. Detailed methodology and calculations with some fiducial examples  are presented.
\end{abstract}

\begin{keywords}
Physical data and process: turbulence -- techniques: interferometric -- ISM: clouds -- method: statistical
\end{keywords}

%%%%%%%%%%%%%%%%%%%%%%%%%%%%%%%%%%%%%%%

\section{Introduction} 
Turbulence plays a significant role to shape the  structure, dynamics and  physical condition \citep{2004ARA&A..42..211E} of the interstellar medium (ISM). Different methods are discussed in literature \citep{2000ApJ...537..720L, 2004ApJ...616..943L, 2016MNRAS.456L.117D} to probe the turbulence induced scale invariant density and velocity fluctuations in the neutral hydrogen gas (henceforth \HI). At large scale the \HI has been probed by 21-cm emission studies at scales of a few kpc \citep{2001ApJ...548..749E, 2013MNRAS.436L..49D}in external spirals to scales of a few to hundred parsecs in the Galaxy \citep{1983A&A...122..282C, 1993MNRAS.262..327G}. These studies reveal a scale invariant ISM structure over several decades of length scale ranges. At smaller scales of a few AU to parsecs \HI 21-cm emission is rather difficult to detect. Absorption of 21-cm lines against strong background source is used to study the scale structures of the cold phase of the ISM. Measurement of \HI optical depth using VLBI or interstellar scintillation along some lines of sights \citep{2001AJ....121.2706F, 2005AJ....130..698B} in the Galaxy gives evidence of scale-invariant structures at scales of a few pc and lower (see the recent review by \citet{2018ARA&A..56..489S}). An alternative approach is to study the variation of optical depth against a few fast moving pulsars, where \citet{1994ApJ...436..144F} find optical depth variation along a few lines of sights  at 5-100 AU scales, whereas other authors like \citet{2003MNRAS.341..941J,  2003ApJ...598L..23S} find no variation along other directions. A still interesting approach is to measure the two-point function of the optical depth variation against extended background sources like radio galaxy or supernovae remnants \citep{2000ApJ...543..227D, 2010MNRAS.404L..45R,  2012ApJ...749..144R,2014MNRAS.442..647D}, where often scale invariant structure of similar characteristics is seen at different directions in the Galaxy.

It is quite debated in the literature \citep{2000MNRAS.317..199D} whether these small-scale fluctuations are a result of physical processes or they are just a few short-lived entities at a particular line of sight. Existence of small scale coherent structure would indicate a significant change in the process of star formation \citep{2014PhR...539...49K}, whereas the short-lived structures may account for the observed gas in the unstable phases of the ISM. % Resolving this puzzle observationally requires two directions, first to observe the two-point function of the opacity fluctuation in the different direction in the Galaxy to find if this small scale CNM structures are plenty. Secondly, it is important to compare the findings at small scales using optical depth studies with the emission observation at large scales where the specific intensity fluctuations and hence density fluctuations of \HI is directly measured. 
In this report, we address the question of estimating the two-point statistics of the \HI column density in the CNM using the absorption studies and compare the existing results with that at the large scales.  We outline a methodology to estimate the two-point correlation function of the optical depth  (section 2),  investigate the relation with the column density fluctuations (section 3) and  check the efficacy of such estimations with a few fiducial models  (section 4) and discuss the importance of our findings (section 5). 

\section{Two point statistics of the optical depth  from observed visibilities}
\subsection{Specific intensity of 21-cm radiation}
The observed specific intensity $I(\vec{\theta}, \nu)$ towards any extended background source can be written from the simple solution of the radiative transfer equation as 
\begin{equation}
I(\vec{\theta},\nu) = I_{c}(\vec{\theta}) e^{-\tau(\vec{\theta},\nu)}.
\end{equation}
where $I_{c}$ and $\tau$ represents the continuum radiation from an extended background source and  the optical depth of \HI 21-cm line along the direction $\vec{\theta}$ respectively. Here we consider observations with a relatively smaller field of view and $\vec{\theta}$, represents angular separation from the centre of the field of view. It is further assumed that  the continuum radiation is not a function of frequency.  Specific intensity from an extended continuum  source can be modelled as 
\begin{equation}
I_{c}(\vec{\theta}) = W(\vec{\theta})[\bar{I_{c}} + \Delta I_{c}(\vec{\theta})],
\end{equation}
were $\bar{I_{c}}$ is the total flux from the background source, the function $W(\vec{\theta})$, termed as the window function, gives the large scale distribution of specific intensity in the sky with $\int d \vec{\theta} W(\vec{\theta}) = 1$. Here we model the small scale fluctuations in the specific intensities with $\Delta I_{c}(\vec{\theta})$, with its mean zero. We write the optical depth function $\tau(\vec{\theta},\nu)$  as 
\begin{equation}
  \tau(\vec{\theta},\nu)=\bar{\tau}(\nu)+\Delta \tau(\vec{\theta},\nu),
  \label{eqn:tau1}
\end{equation}
where $\bar{\tau}(\nu) $,  the spatially  average value of the optical depth, is a function of the observed frequency. The observed specific intensity can be written as
\begin{equation}
I(\vec{\theta},\nu)= W(\vec{\theta}) \left[ \bar{I_{c}} + \Delta I_{c}(\vec{\theta}) \right ] e^{-\left [\bar{\tau}(\nu)+\Delta \tau(\vec{\theta},\nu)\right ] }
\end{equation}
\subsection{Auto-correlation function of optical depth fluctuations}
The two-point correlation or the auto-correlation function of the observed specific intensity function is defined as
\begin{equation}
\xi_{I}(\vec{\theta}, \vec{\theta'}, \nu) = \langle I(\vec{\theta},\nu) I(\vec{\theta '},\nu) \rangle,
\end{equation}
where the angular brackets correspond to ensemble averaging. The two-point correlation function of the specific intensity function of the continuum radiation can be written as 
%\begin{equation}
$\xi_{I_{c}}(\vec{\theta}, \vec{\theta'}) = \langle I_{c}(\vec{\theta}) I_{c}(\vec{\theta '}) \rangle$.
%\end{equation}
Hence,   for the frequencies with absorption the two-point correlation function of the specific intensity fluctuations can be written as (see \citet{2010MNRAS.404L..45R})
\begin{equation}
\xi_{I}(\vec{\theta},\nu)=e^{-2\bar{\tau}(\nu)+\sigma^{2}_{\tau}(\nu)} \xi_{I_{c}}(\vec{\theta}) \  e^{\xi{\tau}(\vec{\theta},\nu)},
\end{equation}
where $\sigma_{\tau}(\nu)$ is the standard deviation of the optical depth in a given frequency over all angular directions. The quantity $\xi_{\tau}$ is the two point auto correlation function of the opacity fluctuation,
\begin{equation}
\xi_{\tau}(\vec{\theta}, \nu) = \langle \Delta \tau(\vec{\theta}, \nu) \Delta \tau(\vec{\theta}', \nu) \rangle.
\end{equation}

\begin{table}
\begin{tabular}{|l|c|c|}
\hline
  & Auto-correlation  & Power spectrum   \\
\hline
Continuum   & $\xi_{I_{c}}(\vec{\theta})$  & $P_{Ic}(\vec{U})$ \\ 
\hline
Line  & $\xi_{I}(\vec{\theta},\nu)$ & $P_{I}(\vec{U},\nu)$ \\ 
\hline
Optical depth & $\xi_{\tau}(\vec{\theta},\nu)$  & $\Xi(\vec{U},\nu)$ \\ 
\hline 
\end{tabular}
\caption{Different two point statistics used in the analysis. Note that $\Xi(\vec{U},\nu)$ is not the direct power spectrum of optical depth and is defined in eqn.~(\ref{eqn:xi}).}
\end{table}
\subsection{Measuring the optical depth autocorrelation from radio interferometric observations}
We use radio interferometers to measure the sky brightness distributions. The measured visibility  $V(\vec{U},\nu)$ at different inverse angular scale or baselines $\vec{U}$ and frequency $\nu$ can be approximated as  Fourier transform \citep{2017isra.book.....T} of the sky intensity fluctuation as
\begin{equation}
V(\vec{U},\nu)=\int A(\vec{\theta}) W(\vec{\theta}) I({\vec{\theta}},\nu)  e^{-2\pi \vec{U}.\vec{\theta}}{d\vec{\theta}},
\end{equation}
where $A(\vec{\theta})$ is the antenna beam pattern.
We use $V_{2}(\vec{U},\vec{U'},\nu)$ to denote visibility correlation  $\langle V(\vec{U},\nu) V^{*}(\vec{U'},\nu)\rangle $, which can be written as  
\begin{equation}
V_{2}(\vec{U},\vec{U'},\nu)= 
 |\tilde{W}(\vec{U})|^{2} \otimes[2\pi\delta(\vec{U}-\vec{U'}) P_{I}(\vec{U},\nu) ],
\end{equation} 
 where   $\otimes$ represents convolution. Here we have used $ P_{I}(\vec{U},\nu)$ to denote  the  power spectrum of the specific intensity, which is related to the autocorrelation function as 
 \begin{equation}
  P_{I}(\vec{U},\nu)=\int \xi_{I}(\vec{\theta}, \nu)\  e^{-2\pi \vec{U}. \vec{\theta}}d\vec{\theta}
 \end{equation}
 and $\tilde{W}(\vec{U})$ is given as
%  \begin{equation}
$\tilde{W}(\vec{U})=\int A(\vec{\theta}) W(\vec{\theta})  e^{-2\pi \vec{U}.\vec{\theta}}{d\vec{\theta}}.$
%\end{equation}
Note that, since the angular extent of the continuum sources of interest ($\le 10^{'}$) are a factor of three or more smaller than the typical field of view of the interferometers (like $\sim 45^{'}$ or even larger for VLA, GMRT), $\tilde{W}(\vec{U})$ will be dominated by the Fourier transform of $W(\vec{\theta})$.
 At the frequencies without \HI absorption,  visibility correlation $V_{2c}(\vec{U},\vec{U'})$ measures the  power spectrum  $P_{Ic}(\vec{U})$ of the continuum specific intensity.
% \begin{equation}
%  V_{2c}(\vec{U},\vec{U'})= |\tilde{W}(\vec{U})|^{2} \otimes[2\pi\delta(\vec{U}-\vec{U'}) P_{\delta Ic}(\vec{U})],
%\end{equation} 
We can write the visibility correlation at the channels with \HI absorption as
\begin{equation}
V_{2}(\vec{U},\vec{U'},\nu)= e^{-2 \bar{\tau}(\nu)+\sigma^{2}_{\tau}(\nu)} V_{2c}(\vec{U},\vec{U'}) \otimes \Xi (\vec{U},\nu),
\end{equation}
where
\begin{equation}
  \Xi(\vec{U}, \nu) = \int d \theta e^{\xi_{\tau}(\vec{\theta}, \nu)} e^{- 2 \pi \vec{U}. \vec{\theta}}.
  \label{eqn:xi}
\end{equation}
The function $\Xi(\vec{U},\nu)$ carries information of the two point statistics of the  opacity fluctuation. In this work we are interested in cold neutral medium structures at length scales ranging sub parsecs to a few tens of parsecs.  Recent works by \citet{2015ApJ...809..153M, 2016A&A...595A..37K, 2017A&A...607A..15K, 2017ApJ...834..126B}  and \citet{2019ApJ...874..171C} find the existence of  anisotropic structures at scales of a few pcs to 100s of pcs in certain directions in the Galaxy with significant alignments along the magnetic field lines.    Here, we restrict ourselves to the isotropic estimators of the two point correlation function and hence  $\Xi$  depends only on $U= |\vec{U}|$. We first discuss the sensitivity of different physical parameters of the ISM  assuming isotropic fluctuations in the column density in section~4.1.  In Section~4.2, we access the signature of anisotropic structures on the isotropic estimator of the ISM structures.
\section{Relation of optical depth autocorrelation with statistics of the \HI column density and spin temperature}
% Plot of Chi with angular scales
\begin{figure}
\begin{center}
\includegraphics[width=0.8\columnwidth]{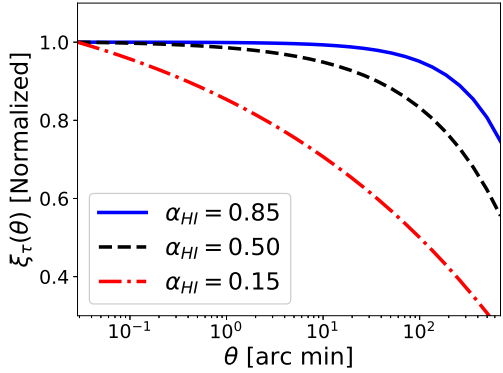}
\caption{Autocorrelation function of optical depth plotted as a function of angular scales for different values of $\alpha_{HI}$.  Value of $\gamma.$ is taken to be zero for all subsequent plots unless stated explicitly.}
\label{fig:xi}
\end{center}
\end{figure}
The optical depth of  \HI 21-cm radiation along the direction $\vec{\theta}$  in an observing frequency $\nu$ can be written as
\begin{equation}
 \tau(\vec{\theta},\nu) = \frac {3 h c^{2} A_{21 }} {32 \pi k_{B} \nu_{0} } \int  {\rm d} z\  \frac {  n_{HI}(\vec{\theta},z) }{ T_{s}(\vec{\theta},z)} \ \phi(\vec{\theta}, z, \nu)    
  \end{equation}
where $A_{21} $,  $k_{B}$ and $\nu_{0}$   are the Einstein  A coefficient, Boltzmann constant and rest frequency of 21-cm emission, $z$ is the line of sight direction, $n_{HI}$ denotes the \HI number density and $T_{s}$ is the spin temperature of the gas. The function $\phi$ is called the line shape function and defined such that $\int \phi(\nu) {\rm d} \nu = 1$.   In principle, the line shape may vary at different directions and line of sight depth. We assume to observe with a frequency width such that the line shape function can be considered as constant. We write the number density of the optical depth fluctuation and the spin temperature in terms of their corresponding mean (denoted by bar `` $\bar {}$ ") as
\begin{eqnarray}
n_{HI}(\vec{\theta}, z) = \bar{n}_{HI} \left[ 1 + \delta _{n_{HI}} (\vec{\theta}, z) \right ] \\ \nonumber
 T_{s}(\vec{\theta}, z) = \bar{T}_{s} \left[ 1 + \delta _{T_{s}} (\vec{\theta}, z) \right ].
\end{eqnarray}
Assuming that the fluctuations to be  small compared to unity, we may write to the first order 
\begin{equation}
  \tau(\vec{\theta},\nu) =   \bar{\tau} (\nu) \int \frac{{\rm d} z}{L} \left [ 1 + \delta n_{HI}  (\vec{\theta},z) - \delta_{ T_{s}}(\vec{\theta},z)  \right],
  \label{eqn:tau2}
  \end{equation}
  where we have identified the average optical depth as 
%  \begin{equation}
 $ \bar{\tau}(\nu) =
 \frac {3 h c^{2} A_{21 }} {32 \pi k_{B} \nu_{0} } \frac{\bar{N}_{HI}}{\bar{T}_{s}} \phi(\nu),$
% \end{equation}
 where  $L$ is the characteristic line of sight depth of the cloud and $ \bar{N}_{HI} = \int {\rm d} z \bar{n}_{HI}$ is the spatially  averaged column density. Henceforth,  we shall consider all quantities at the single frequencies only and will not write the frequency dependence explicitly. Using eqn.~(\ref{eqn:tau1}) and  eqn.~(\ref{eqn:tau2}) the optical depth fluctuation at different directions in the sky can be written as
 \begin{equation}
 \Delta \tau(\vec{\theta}) = \frac{\bar{\tau}}{L} \int {\rm d} z \left [ \delta n_{HI}  (\vec{\theta},z) - \delta_{ T_{s}}(\vec{\theta},z)  \right].
 \end{equation}
 Assuming homogeneity and isotropy in the column density and spin temperature distribution the autocorrelation function of optical depth fluctuation can be written as
 \begin{equation}
 \xi_{\tau}(\theta) = \frac{\bar{\tau}^2}{L}  \int {\rm d} z  \left [  \zeta_{HI}(\theta, z)  +  \zeta_{T_{s}}(\theta, z)-  2 \zeta_{HT}(\theta, z)  \right],
  \end{equation}
 with the autocorrelations and cross correlations of \HI number density and spin temperature being defined like
 \begin{equation}
\zeta_{ab}(\theta, z) = \langle \delta a(\vec{\theta}, z) \delta b(\vec{\theta}', z') \rangle .
\end{equation}
To proceed further,  we need to consider the thermodynamic properties of the CNM cloud.  If  we consider that the clouds are   in both thermal and pressure equilibrium, the fluctuations in the spin temperature will be directly related to the column density, i.e $\delta T_{s} = - \delta n_{HI}$. On the other hand, it is possible that the CNM clouds are in thermal equilibrium and behaves like adiabatic gas, then $\delta T_{s} = - \delta n_{HI} (1 - \gamma)$, where $\gamma=5/3$ is the adiabatic index.  In addition to this, the CNM condition can be with no correlation between  the  column density and  the spin temperature. In general  a combination of correlated and uncorrelated gas may  coexist, i.e $\delta T_{s} = - \eta ( 1- \gamma)\, \delta n_{HI} + ( 1 - \eta) \times $ uncorrelated part, where $\eta$ gives the fractional correlation.  The autocorrelation can be written as 
\begin{equation}
 \xi_{\tau}(\theta) = \bar{\tau} ^{2}  \left [  \chi \  \xi_{HI}(\theta)  + ( 1 - \eta )^{2} \sigma_{T_{s}^{2}} \right],
\end{equation}
 when $\chi = \left[1 + \eta (1-\gamma)\right]^2$ and $\xi_{HI}(\theta) = \frac{1}{L}\, \int {\rm d} z \zeta_{HI}(\theta, z)$.
 Since autocorrelation at $\theta=0$ gives the corresponding variance, we may write
 \begin{equation}
 \sigma_{\tau}^2 =  \bar{\tau} ^{2}  \left [  \chi \ \sigma_{HI}^2  + (1 - \eta)^{2} \sigma_{T_{s}^{2}} \right],
 \end{equation}
which implies, for a scale invariance column density fluctuation
\begin{equation}
  \xi_{\tau}(\theta) = \sigma^{2}_{\tau} - \chi\ \bar{\tau}^{2} \sigma_{HI}^{2} \theta^{\alpha_{HI}},
  \label{eq:final}
\end{equation}
for a power law opacity fluctuation power spectrum with power law index $-2-\alpha_{HI}$.

Note that the global statistics of the optical depth ($\bar{\tau}, \sigma_{\tau}$ etc) can be directly estimated from the optical depth map itself \citep{2014MNRAS.442..647D, 2019RAA....19...60D}. The quantity $\Xi(\vec{U}) $ carries additional information about the parameters $\sigma_{HI}, \eta$, $\gamma$ and  $\alpha_{HI}$. However, it is clear from eqn.~(\ref{eq:final}) that there is a degeneracy in estimates of $\eta$, $\gamma$ and $\sigma_{HI}$ if only the optical depth autocorrelation function is used as a measure. 

\section{Sensitivity of observables on the physical parameters  }
\subsection{With isotropic column density fluctuations}
\begin{figure*}
  \begin{center}
    \includegraphics[width=1.9\columnwidth]{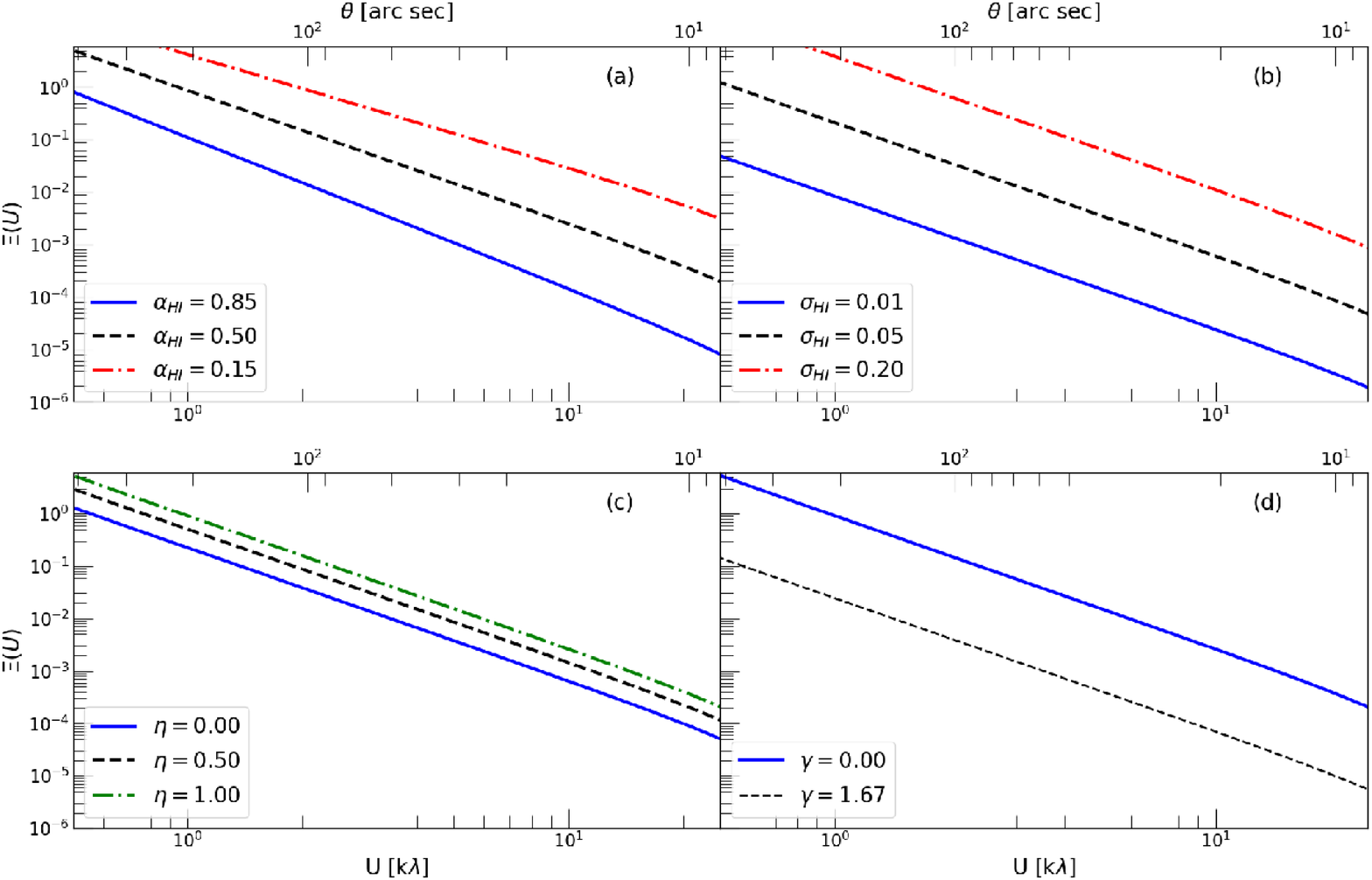}
    \caption{Variation of $\Xi(U)$ with respect to baseline $U$ for different values of (a) $\alpha_{HI}$, (b) $\sigma_{HI}$, (c) $\eta$ and (d)$\gamma$.}
    \label{fig:zeta}
  \end{center}
\end{figure*}
 
% Plot of CHI with  baseline: Varying alp_HI(a), std_HI(b), eta_HI(c)
% and gamma(d)
%\begin{figure*}
%  \begin{center}
%    \subfloat{\includegraphics[width=0.75\columnwidth]{Figure_CHI_alpHI.eps}}
%    \subfloat{\includegraphics[width=0.75\columnwidth]{Figure_CHI_stdHI.eps}}\\
%    \subfloat{\includegraphics[width=0.75\columnwidth]{Figure_CHI_Chi.eps}}
%    \subfloat{\includegraphics[width=0.75\columnwidth]{Figure_CHI_gamma.eps}}

%    \caption{Variation of $\Xi(U)$ with respect to baseline $U$ for different values of (a) $\alpha_{HI}$, (b) $\sigma_{HI}$, (c) $\eta$ and (d)$\gamma$.}
%    \label{fig:zeta}
%  \end{center}
%\end{figure*}

In this section, we investigate sensitivity of $\Xi(U)$ on the parameters $\sigma_{HI}, \eta$, $\gamma$ and $\alpha_{HI}$ assuming the fluctuations in the column density to be isotropic. In a typical observation of \HI in absorption we wish to estimate the quantities $\bar{\tau}$ and $\sigma_{\tau}$ from the reconstructed images.  Following the calculations of \citet{2014MNRAS.442..647D}  the bias in estimating these quantities are lowest at an optical depth of unity. Hence, in this study we choose our fiducial value for the average optical depth to be $1$ with a standard deviation $\sigma_{\tau} = 0.1$. In order to unbiasedly estimate the scale-dependent two-point statistics of $\tau$, we use visibility correlations.

We consider two extreme values of the parameter $\gamma$, i.e $\gamma = [0, 5/3]$.  Note that, with $\gamma=0$, $\chi$ have a value of unity for $\eta = 0$ and a maximum value of four for $\eta = 1$. With $\gamma = 5/3$, $\chi$ varies rather slowly as $\eta$ is changed from $0$ to $1$. To investigate the effect of $\eta$ on the measured  $\Xi(\vec{U})$, we choose three  values for $\eta$,  $[0., 0.5, 1.0]$ with the fiducial value being $\eta = 0.5$. \HI emission studies at large scale suggest that the fluctuations in the  density of \HI is about $10$ $\%$ of the mean \citep{2013MNRAS.436L..49D},  hence we choose three values of $\sigma_{HI}$ as $[0.01, 0.05, 0.20]$ to see the variation against it, the fiducial value of $\sigma_{HI}$ is chosen to be $0.1$. \citet{2010MNRAS.404L..45R} measure the opacity fluctuation power spectrum of the \HI along the direction of the supernovae remnant Cassiopea A. Their findings are consistent with a model when the standard deviation of the optical depth varies with scales as $\sigma_{\tau} = 2 \left( x / 4 pc \right)^{0.34}$. This would correspond to the  slope of $0.68$ for the $\alpha_{HI}$. Hence, to see the variation of the $\Xi(\vec{U})$  we consider three values of $\alpha_{HI}$, $[0.15, 0.5, 0.85]$ with the fiducial value at $0.5$.

We start by generating a  zero mean Gaussian random field corresponding to the \HI optical depth  fluctuation given by the  autocorrelation function in eqn.~(\ref{eq:final}). We then calculate the quantity $\Xi(U)$ of this field using two dimensional Fast Fourier transform as in eqn.~(\ref{eqn:xi}).   We examine the shape, slope and amplitude of the function $\Xi(U)$  for the dependence of various physical parameters.

Figure~(\ref{fig:xi}) show the normalised autocorrelation function of the optical depth fluctuations for three values of $\alpha_{HI}$.  Here, we have chosen $\gamma = 0, \eta =0.5$ and $\sigma_{HI} = 0.1$.  Four panels of Figure~4 show the variation of $\Xi(U)$ with baseline for different sets of parameters. The range of angular scales plotted here are from $7 -700$ arcsec, which corresponds to typical baseline ranges ($\sim 0.5 - \sim 20$ k$\lambda$) such observations will be sensitive to. 

In Figure~\ref{fig:zeta}(a) we plot the function $\Xi(U)$ as a function of  baseline $U$ corresponding to the autocorrelation in Figure~(\ref{fig:xi}). The function  $\Xi(U)$ assumes power law with the power law index as $[-2.85, -2.5, -2.15]$ respectively for  $\alpha_{HI} = [0.85, 0.5,0.15]$. Since we consider small fluctuations of optical depth over the mean, the quantity $\Xi(U)$ is expected to have the same functional form as the power spectrum of the optical depth itself, i.e the slope of  $\Xi(U)$ as a function of $U$ when plotted in log-log scale is expected to be $-2 - \alpha_{HI}$ (see discussion in \citet{1975ApJ...202..439L}). In this view, the relation between the slope of $\Xi(U)$ and $\alpha_{HI}$ is as expected.  This demonstrates that, for scale independent fluctuation in the \HI column density, irrespective of the values of the parameters $\gamma, \eta$ and $\sigma_{HI}$  we can estimate the power law index of the \HI column density fluctuation from observations of optical depth fluctuations.

The amplitude of the function $\Xi(U)$ is sensitive to all three of the parameters $\gamma, \eta$ and $\sigma_{HI}$. Fixing $\gamma= 0$, we first show the variation of $\Xi(U)$ with different values of $\sigma_{HI}$ in Figure~\ref{fig:zeta}(b). Note that, the power spectrum of a fluctuation is directly proportional to the square of its standard deviation. Since for small fluctuations in optical depth, $\Xi(U)$ has the same functional form as the power spectrum of optical depth, the  amplitude of $\Xi(U)$ is seen to be proportional to $\sigma_{HI}^2$. We show the variation of $\Xi(U)$ with different values of  $\eta$ in Figure~\ref{fig:zeta}(c).  Finally, we fix all but $\gamma$ to the  fiducial values and plot $\Xi(U)$ for two extreme cases of $\gamma = 0$ and $5/3$ in Figure~\ref{fig:zeta}(d). In practice, without any prior information of any two of $\gamma, \eta$ and $\sigma_{HI}$  it will not be possible to estimate these parameters separately only from the  observation of optical depth fluctuations.

\subsection{Effect of anisotropic fluctuations}
In this work we use an isotropic estimator of the optical depth fluctuation and relate it to the isotropic statistics of the column density fluctuations. \citet{1995ApJ...438..763G} discuss a model of ISM structure shaped by  Magnetohydrodynamic (MHD) turbulence. Such MHD turbulence is expected to produce scale independent anisotropic structures with the degree of anisotropy governed mostly by the Alfvenic Mach number \citep{2011ApJ...740..117E, 2014ApJ...790..130B}. In a  more recent study on the anisotropy in ISM turbulence, \citet{2017MNRAS.464.3617K}  quantify the anisotropic structures in terms of the multipole moments of the structure function.  \citet{2004ApJ...616..845M}, \citet{2016A&A...595A..37K} and \citet{2017A&A...607A..15K} measure the anisotropy associated with \HI distribution along different directions in our Galaxy and the Magellanic bridge. 

All these studies  observe that the anisotropic structure in \HI is mostly scale independent,  as suggested by the theoretical works cited above. They quantify the power anisotropy as $Q = P(U, \phi=0)/P(U, \phi=\pi/2)$, where $P(U, \phi)$ is the observed power spectrum as a function of the baselines and azimuthal angle $\phi$ with $\phi=0^{\circ}$ corresponding to the azimuthal direction with maximum power.  

As discussed in the previous section, the directly measured quantity $\Xi(U)$, we use in this work to estimate the fluctuations in the column density,   is proportional to the  optical depth power spectrum for small optical depth fluctuations. In this section we investigate how the amplitude of an  isotropic estimator of the power spectrum $P(U)$ and hence $\Xi(U)$  is modified in presence of anisotropic structures in the \HI. We start by modelling scale independent anisotropic power spectra of \HI optical depth, i.e,  $P_{\tau} (\vec{U})$ in polar coordinate $(U, \phi)$, as 
\begin{eqnarray}
P_{\tau}(U, \phi) &=& A\  \Phi(\phi) \ U^{-2-\alpha_{NHI}},\ \ {\rm where}\\ \nonumber
\Phi(\phi) &=& \left [ \frac{f_a \cos^2 \phi + 1}{f_a + 1}\right].
\label{eq:anisops}
\end{eqnarray}
Here, the overall amplitude of the power spectrum $A$ gives the isotropic power. The anisotropy in the power spectrum is incorporated in the function $\Phi(\phi)$, with the quantity $f_a$ as the anisotropy parameter. The factor $f_a$ is related to the  parameter $Q$ used in literature as $Q = f_a + 1$. In this model, $f_a=0$ corresponds to isotropic fluctuations with $Q=1$.  
\begin{figure}
  \begin{center}
    \includegraphics[width=.8\columnwidth]{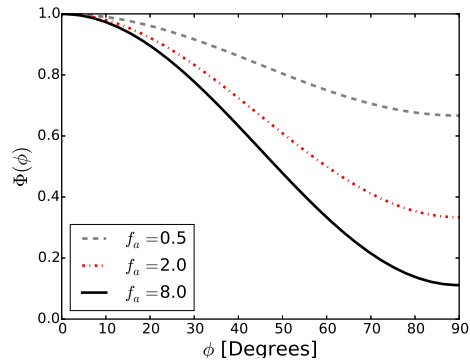}
    \caption{Variation of $\Phi(\phi)$ at different directions in the baseline plane is shown  for different values of the anisotropy factor $f_a$.}
    \label{fig:anisomodel}
  \end{center}
\end{figure}
 Figure~\ref{fig:anisomodel} shows the function $\Phi(\phi)$ for different values of $f_a$.  

If we use an isotropic power spectrum estimator  $P(U)$ to quantify the fluctuations in the optical depth given by the anisotropic power spectrum above, then, 
\begin{eqnarray}
P(U) &=& \frac{1}{2\ \pi} \int_0 ^{2 \pi} P(U, \phi) \ {\rm d} \phi \\ \nonumber 
&=& A\ \mathcal{R}(f_a)\  U^{-2-\alpha_{NHI}},\ \ {\rm where} \\ \nonumber
\mathcal{R}(f_a) &=& \frac{f_a + 2}{2 ( f_a + 1 )}.
\end{eqnarray}
 \begin{figure}
  \begin{center}
    \includegraphics[width=.8\columnwidth]{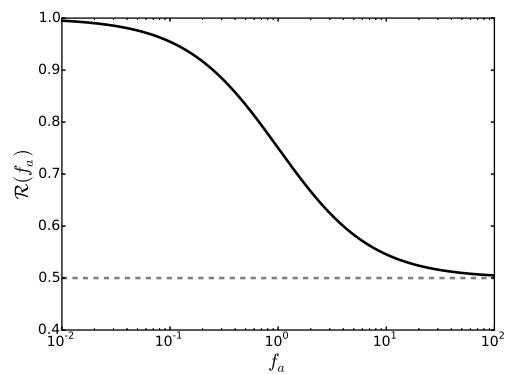}
    \caption{Variation of $\mathcal{R}$   for different values of the anisotropy factor $f_a$ (in dark black). The dashed grey line shows the asymptotic value $1/2$.}
    \label{fig:anisopower}
  \end{center}
\end{figure}
Hence, the effect of using an isotropic power spectrum estimator to quantify anisotropic  and scale invariant structures in \HI results in a scaled estimate of the amplitude of the power spectra $A$. We conclude that the presence of anisotropic fluctuations will result in an additional factor of $\mathcal{R}(f_a)$ in the amplitude of $\Xi(U)$ .  Note that, the exact form of this modification depends on the functional form of $\Phi(\phi)$, the result shown here is for the model considered in this work. Figure~\ref{fig:anisopower} shows the variation of the anisotropic power $\mathcal{R}(f_a)$with the anisotropy parameter $f_a$ for our model. Interestingly, for high values of anisotropy, the correction factor $\mathcal{R}$ approaches $1/2$ in this model.
 
\section{Discussion and Conclusion}
In this article, we investigated the possibility of estimating the physical parameters of the CNM clouds from observation of optical depth fluctuation against an extended background source. It is expected that the compressible fluid turbulence would create scale invariant column density fluctuation in the CNM cloud. Our investigation shows, such observations alone can not trace much on the amplitude of such fluctuations. In fact, the partial correlation between spin temperature and column density, the dependence of spin temperature on the column density in the correlated case and fluctuation amplitude of \HI column density all produce a similar effect on the measured optical depth autocorrelation. Multi-wavelength observations (see \citet{2011ApJ...728..159W}) along the same line of sight can break this degeneracy, whereas using a different spectral line we may arrive at the spin temperature fluctuation  independently, if the assumption of thermal equilibrium is valid. Also, observation of recombination lines of different species with high dynamic range spectral calibration (see \citet{2019MNRAS.486...42C})  may independently hint on the statistics of fluctuation of spin temperature and resolve the degeneracy discussed above. On the brighter side, we find that the scale dependence of such density fluctuations at sub-parsec  scales can be traced by the radio-interferometric observations of the optical depth fluctuations.

Using numerical simulations of $^{13}$CO specific intensity, \citet{2013ApJ...771..123B} show that the slope of the intensity fluctuation power spectra for optical thick clouds saturates at $-3.0$, as expected from the prediction of \citet{2004ApJ...616..943L}, whereas the spectral slope of the optically thin clouds is shallower. Recent simulations of galactic ISM by \citet{2010MNRAS.409.1088B} with an LMC size model find that the power spectrum of gas surface density assumes a slope of $\sim-2.8.$ to $-3.0$ at scales $100$ pc and lower depending on the dynamics is driven by only gravity or additionally by star formation feedback. \citet{2017MNRAS.466.1093G} show that for Milky Way size galaxies, the effect of stellar feedback produces a slope of $-2.5$ for the \HI column density power spectra at scales of $1$ kpc or smaller. Observations of the \HI structures are done in emission for  typical  scales of $\sim 1$ pc and higher, where the column density power spectrum is directly measured. At lower scales, the \HI structures are traced by the observed structure in optical depths. \citet{1983A&A...122..282C} report measurement of \HI intensity fluctuation power spectrum along the galactic plane with a power law slope varying between $-2.0$ to $-3.0$. In a subsequent study,  \citet{1993MNRAS.262..327G} measures the power spectrum slope to be between $-2.2$ to $-3.0$ in different regions in the Galaxy.  
\citet{2010MNRAS.404L..45R} report observation of the optical depth power spectrum with the slope of $-2.86$ from the local arm towards the direction of Cassiopeia A. On the other hand, \citet{2000ApJ...543..227D} find the slope of $ - 2.5 $ towards the Cygnus A from the local arm and  $ -2.75 $  towards the Perseus arm. \citet{2018ApJ...856..136P} use different tracers including \HI to estimate the power spectrum towards the Perseus molecular cloud and report the slope of $-3.23$ for their opacity corrected power spectra.  \citet{2013NewA...19...89D} measure \HI power spectra for 16 spiral galaxies from the THINGS \citep{2008AJ....136.2563W} sample. They find the power spectra of column density fluctuations follow a power law with slope mostly in the range $-1.5$ to $-1.8$ at the scales of $\sim1 -10$ kpc for the 2D structures in \HI column density. If the same turbulence cascade is continued, these correspond to the  slope of $-2.5$ to $-2.8$ at the  sub pc scales \citep{2009MNRAS.397L..60D}, owing to a geometrical effect.  On the other extreme of the length scales, \citet{2012ApJ...749..144R, 2014MNRAS.442..647D} observe that the power spectrum of \HI optical depth fluctuations at scales of  few AU has a power law of $-2.81$. More recently, \citet{2016A&A...593A...4M} report that the power spectrum of dust along the direction of cirrus follow a power law of $-2.9$ down to a scale of $0.01$ pc. These observations along different directions in the Galaxy and at different length scales raise a couple of questions, (i) is there a singular turbulence cascade responsible for ISM turbulence at scales from $10$ kpc to $1$ mili pc, (ii) or if it is just a mere coincidence that we are only sampling the small scale structures in the Galaxy towards a few directions. We have identified about 15 supernovae remnants which can be used to scan the galactic plane at different latitude and a few lines of sights have already been observed and analysed.

In this work, we studied the effect of anisotropic structures in \HI  in the isotropic power spectrum using a particular model for the anisotropic power spectrum. \citet{2016A&A...595A..37K} and \citet{2017A&A...607A..15K} measure a maximum power anisotropy of a few times $10$, which translate to a factor of $\sim0.5$ for the quantity $\mathcal{R}$. This suggests that in  presence of anisotropic structures in \HI, an estimation of only isotropic power results in an additional degeneracy in the amplitude of the measured optical depth autocorrelation. This degeneracy, however, can be lifted by probing the statistical nature of anisotropic structures. Though this  requires considerably large observation time, it can be pursued in principle and it would provide more insight on  nature and generation mechanisms of the ISM structures.  
 
 Here, we have taken the \HI clouds to have three dimensional geometry and have not considered averaging over velocity space as discussed in \citet{2000ApJ...537..720L} to estimate the statistics of velocity fluctuations. In case of thin clouds, we expect the power law index of $\zeta$ to be one lower than $\xi$ (see \citet{2009MNRAS.397L..60D}) and column density power spectra can be estimated from the autocorrelation of the optical depth. On the other hand, though velocity channel analysis \citep{2004ApJ...616..943L} would have given more information about the energetics of the CNM turbulence at these scales, considering thicker velocity slices breaks down our assumption of uniform line shape function. Relating the autocorrelation  function of optical depth with thicker velocity slices to \HI column density power spectrum requires more investigation and not  in the scope of this paper.  In a nutshell, this work demonstrates how radio-interferometric spectral line absorption studies can be used to estimate the power spectrum of the \HI column density fluctuation. 
%We expect to use this technique in a forthcoming paper and present  results from observations along a few lines of sights. 

\section*{Acknowledgement}
PD would like to acknowledge support from DST-INSPIRE fellowship for this work.

\bibliographystyle{mnras}
\bibliography{ref} 

% Don't change these lines
\bsp	% typesetting comment
\label{lastpage}
\end{document}